\documentclass[preprint,showpacs,amssymb,amsmath,aps,pra]{revtex4}
\newcommand{\half}{\mbox{$\frac{1}{2}$}}
\usepackage{graphics}
\usepackage{bm}
\begin{document}
\title{Description of thermal entanglement with the static path plus random
 phase approximation}
\author{N. Canosa, J.M. Matera, R. Rossignoli}
\affiliation{Departamento de F\'{\i}sica-IFLP,
Universidad Nacional de La Plata, C.C.67, La Plata (1900), Argentina}
\begin{abstract}
We discuss the application of the static path plus random phase approximation
(SPA+RPA) and the ensuing mean field+RPA treatment to the evaluation of
entanglement in composite quantum systems at finite temperature. These methods
involve just local diagonalizations and the determination of the generalized
collective vibrational frequencies. As illustration, we evaluate the pairwise
entanglement in a fully connected XXZ chain of $n$ spins at finite temperature
in a transverse magnetic field $b$. It is shown that already the mean field+RPA
provides an accurate analytic description of the concurrence below the mean
field critical region ($|b|<b_c$), exact for large $n$, whereas the full
SPA+RPA is able to improve results for finite systems in the critical region.
It is proved as well that for $T>0$ weak entanglement also arises when the
ground state is separable ($|b|>b_c$), with the limit temperature for pairwise
entanglement exhibiting quite distinct regimes for $|b|<b_c$ and $|b|>b_c$.
\end{abstract}
\maketitle

\section{Introduction}
It is now well recognized that quantum entanglement plays an essential role in
both quantum information science \cite{NC.00}, where it is considered a {\it
resource}, as well as in many-body and condensed matter physics, where it
provides a new perspective for understanding quantum correlations and critical
phenomena \cite{ON.02,OS.02,V.03,T.04}. Entanglement denotes those correlations
with no classical analogue that can be exhibited by composite quantum systems,
which constitute, for instance, the key ingredient in quantum teleportation
\cite{Be.93}. A pure state of a composite system is entangled if it is not a
product state, while a mixed state of such system is entangled when it cannot
be written as a convex combination of product states \cite{W.89}.

Thermal entanglement \cite{ON.02,ABV.01,Wa.02,W.04,AK.04,CR.04} denotes that of
mixed states of the form $\rho(T)\propto \exp[-\beta H]$, where $H$ is the
system Hamiltonian and $\beta=1/kT$ the inverse temperature. A complete
characterization of thermal entanglement in many component systems is
difficult, since, to begin with, there is no simple necessary and sufficient
computable criterion for determining if a general mixed state is entangled
\cite{DPS.04}. Besides, these systems exhibit entanglement at different levels,
i.e., between any pair or set of subsystems, starting from that between
elementary constituents $i,j$ and ending in that of global partitions
\cite{RC.05} (which for $T>0$ can no longer be measured through the entropy of
a subsystem). Finally, a basic difficulty is the accurate evaluation of
$\rho(T)$ and the ensuing reduced densities $\rho_{ij}$. Standard methods like
the mean field approximation (MFA), which may provide a correct basic
description of thermodynamic observables in some systems, are not suitable for
the evaluation of entanglement since they are based on separable
(non-entangled) trial densities. In small finite systems fluctuations of the
order parameters become important \cite{S.72} and the MFA is to be replaced at
least with some average over different mean field densities, but such an
approach will still fail to describe entanglement as it is essentially based on
a convex combination of product densities.

The principal goal of this work is to show the applicability of the static path
plus random phase approximation (SPA+RPA) \cite{PBB.91,AA.97,RC.97,RCR.98} to
the determination of thermal entanglement. The approach is derived from the
path integral representation of the partition function obtained with the
Hubbard-Stratonovich transformation \cite{HS.57}, and has been applied to the
description of basic observables in diverse fermionic models of nuclear and
condensed matter physics \cite{PBB.91,AA.97,RC.97,RCR.98,CR.01,KK.05}. It takes
into account both the large amplitude static fluctuations (SPA) of the mean
field order parameters, essential in critical regions of finite systems,
together with small amplitude quantum fluctuations (RPA), which may account for
most quantum effects if $T$ is not too low and will be responsible for
entanglement. It also provides a fully consistent MFA+RPA approach
\cite{RCR.98}, obtained through the saddle point approximation to the full
treatment. Here we will formulate the method for a system of $n$
distinguishable constituents, where it involves in principle just local
diagonalizations.

We will employ the formalism to evaluate the thermal pairwise entanglement in a
fully connected $XXZ$ chain of $n$ qubits or spins in the presence of a uniform
transverse magnetic field $b$. Spin chains constitute an attractive scalable
qubit representation for exploring and implementing quantum information
processes \cite{DV.00,DV.98,BB.04} and can be realized in diverse physical
systems, including those based on quantum dots electron-spins \cite{I.99} and
Josephson junction arrays \cite{MS.01}, where the effective model includes
coupling between any two spins. Fully connected symmetric spin models (simplex)
have also intrinsic interest, providing a solvable scenario for examining
entanglement in systems undergoing phase transitions. In particular,
entanglement properties of the fully connected $XX$ and $XY$ model at $T=0$
were thoroughly analyzed in \cite{VPM.04,V.04,V.06}. We will show that the XXZ
model exhibits an interesting non trivial behavior at finite temperature, whose
main features can be correctly described by the SPA+RPA for moderate finite $n$
and even by the MFA+RPA below the critical region, the latter providing an
analytic description which becomes exact for large $n$. The formalism is
described in section II while application to the model is discussed in III.
Finally, conclusions are drawn in IV.

\section{Formalism}
We will consider a composite system described by a Hamiltonian of the form
\begin{equation}
 H=H^0-\half\sum_\nu v_\nu (Q^{\nu})^2\label{Hx}\,,
\end{equation}
where $H^0$, $Q^\nu$ are linear combinations of local operators, i.e.,
$H^0=\sum_i H^0_i$, $Q^\nu=\sum_i Q^\nu_i$, with $H^0_i$, $Q^\nu_i$ acting just
on subsystem $i$ ($Q^\nu_i\equiv I_1\otimes\ldots\otimes Q_{i}^\nu\otimes
\ldots\otimes I_n$, with $[Q^\nu_i,Q^\nu_j]=0$ if $i\neq j$). In a spin chain
$Q^\nu$ could stand, for instance, for total spin operators or general linear
combinations $\sum_i\alpha_i s^{\nu}_i$ of the individual spins $s^\nu_i$.
 {\it Any} quadratic interaction between subsystems,
\begin{equation}
V=-\half\sum_{i,j,\nu,\nu'}O^\nu_i v^{ij}_{\nu\nu'}O^{\nu'}_j\,,
\end{equation}
where $O^\nu_i$ denote local operators, can be written in the diagonal form
(\ref{Hx}) (non-unique), after completing squares or diagonalizing the matrix
$v_{i\nu,j\nu'}\equiv v^{ij}_{\nu\nu'}$, with $Q^\nu$ suitable linear
combinations of the $O^{\nu}_i$. We may assume $v_\nu>0$ in (\ref{Hx}) without
loss of generality if antihermitian operators $Q^\nu$ are allowed
($v_\nu(Q^\nu)^2\rightarrow -v_\nu(iQ^\nu)^2$). In what follows we will
consider finite Hilbert space dimension.

The Hubbard-Stratonovich transformation allows then to express the partition
function $Z={\rm Tr}\exp[-\beta H]$ as the path integral \cite{HS.57}
\begin{eqnarray}
Z&=&\int\! D[x]\,{\rm Tr}\,\hat{T}\exp\{-\!\int_0^\beta\!\!d\tau
[\sum_\nu\frac{x_\nu^2(\tau)}{2v_\nu}+h[x(\tau)]\}\,,\label{Zhs}\\
h(x)&=&\sum_i h_i(x),\;\;\;h_i(x)=H^0_i-\sum_\nu x_\nu Q^\nu_i\,,
\end{eqnarray}
where $\hat{T}$ denotes time ordering and the normalization
$\int\!D[x]\exp[-\!\int_{0}^\beta\!d\tau\!\sum_\nu\!x_\nu^2(\tau)/2v_\nu]=1$ is
assumed. The integrand in (\ref{Zhs}) is essentially the trace of the imaginary
time evolution operator $U[x]$ associated with the path $x(\tau)$ and the
linearized Hamiltonian $h[x(\tau)]$, and is here a {\it product operator}
$\prod_i U_i[x]$, not necessarily positive. Eq.\ (\ref{Zhs}) can be evaluated
by means of a Fourier expansion
\begin{equation}
x_\nu(\tau)=x_\nu+\sum_{n\neq 0} x_{\nu}^{n} e^{i\omega_n\tau}\,,
\;\;\omega_n=2\pi n/\beta\,,
\end{equation}
where $x_\nu\equiv x_\nu^0$ are the {\it static} coefficients, representing the
time average $\langle x_\nu(\tau)\rangle_{[0,\beta]}$, with
$D[x]\propto \prod_{\nu}dx_\nu\prod_{n\neq 0}dx_\nu^n$.

In the SPA+RPA \cite{PBB.91,AA.97,RC.97,RCR.98} (to be denoted for brevity as
CSPA (correlated SPA)), the integrals over the static coefficients $x_\nu$ are
fully preserved, while those over $x_{\nu}^n$, $n\neq 0$, are evaluated in the
{\it saddle point approximation} for each value of the $x_\nu$. The aim is to
take into account large amplitude static fluctuations, which are particularly
relevant in the transitional regions of finite systems, together with small
amplitude quantum fluctuations, which should in principle account for most
quantum effects if the temperature is not too low. The final result can be
expressed as \cite{RCR.98}
\begin{eqnarray}
Z_{\rm CSPA}&=&\int_{-\infty}^\infty e^{-\beta \sum_\nu x_\nu^2/2v_\nu}Z(x)
C_{\rm RPA}(x)d(x)\,,\label{Zcspax}\end{eqnarray}
where $d(x)=\prod_\nu \sqrt{\beta/(2\pi v_\nu)}dx_\nu$ and
\begin{eqnarray}
Z(x)&=&{\rm Tr}\exp[-\beta h(x)]=\prod_i{\rm tr}\exp[-\beta h_i(x)]\,,
\label{Zcspax2}\\
C_{\rm RPA}(x)&=&\prod_{n=1}^\infty{\rm Det}
[\delta_{\nu\nu'}+v_\nu R_{\nu\nu'}(x,i\omega_n)]^{-1}\,,\label{Crpax}\\
R_{\nu\nu'}(x,\omega)&=&\sum_{i,k\neq {k'}}\frac{\langle k_i|Q^\nu_i|{k'}_i
\rangle\langle {k'}_i|Q^{\nu'}_i|k_i\rangle(p_{k_i}-p_{k'_i})}
{\varepsilon_{k_i}-\varepsilon_{k'_i}+\omega}\,. \label{rnux}\end{eqnarray}
with ${\rm tr}$ the local trace,  $|k_i\rangle$ the running local eigenstates
($h_i(x)|k_i\rangle=\varepsilon_{k_i}|k_i\rangle$) and $p_{k_i}=e^{-\beta
\varepsilon_{k_i}}/{\rm tr}\,e^{-\beta h_i(x)}$. Eqs.\
(\ref{Zcspax})-(\ref{rnux}) involve just {\it local} diagonalizations. Eq.\
(\ref{Crpax}) is the RPA correction, fundamental in the present context, which
can be further expressed as \cite{AA.97,RC.97}
 \begin{equation}
C_{\rm RPA}(x)=\prod_{\alpha>0}\prod_{n=1}^\infty
\frac{\lambda_\alpha^2+\omega_n^2}{\omega_\alpha^2+\omega_n^2}=
\prod_{\alpha>0}\frac{\omega_\alpha\sinh(\beta\lambda_\alpha/2)}
{\lambda_\alpha\sinh(\beta\omega_\alpha/2)}\,,
 \end{equation}
where $\alpha\equiv (k_i,k'_i)$ runs over all pairs $k_i\neq k'_i$ ($\alpha>0$
indicating $k_i>k'_i$), $\lambda_\alpha\equiv \varepsilon_{k_i}-
\varepsilon_{k'_i}$ and $\omega_\alpha$ are the running RPA energies,
determined as the roots of the equation
\begin{equation}{\rm Det}[\delta_{\nu\nu'}+v_\nu R_{\nu\nu'}(x,\omega)]=0\,.
\label{w1}\end{equation}
They come in pairs of opposite sign and  can also be obtained as the
eigenvalues of the matrix
\begin{equation}
A_{\alpha\alpha'}(x)=\lambda_{\alpha}\delta_{\alpha\alpha'}+ p_\alpha\sum_\nu
v_\nu Q^\nu_{-\alpha} Q^\nu_{\alpha'}\,, \label{A}
\end{equation}
where $p_\alpha\equiv p_{k_i}-p_{k'_i}$, $Q^\nu_\alpha\equiv\langle
k_i|Q^\nu_i|k'_i\rangle$. Eq.\ (\ref{Zcspax}) can be applied provided $C_{\rm
RPA}(x)>0$, which implies $\omega_\alpha^2+\omega_1^2>0$ $\forall \,\alpha,x$.
Since the lowest RPA energies $\omega_\alpha$ may become imaginary or complex
for $x$ away from the stable mean field solution (see below), the previous
condition sets up a breakdown temperature $T^*$, normally low, such that Eq.\
(\ref{Zcspax}) is applicable for $T>T^*$. Setting $C_{\rm RPA}(x)=1$ in
(\ref{Zcspax}) leads to the plain SPA \cite{S.72}, which, although
significantly improving the MFA in critical regions, is unable to describe
entanglement, as it averages correspond essentially to those of a convex
combination of separable densities (for $h(x)$ hermitian).

{\it MFA+RPA}. Away from critical regions, we may also apply the saddle point
approximation to the static variables $x_\nu$. This leads to the MFA+RPA (to be
denoted as CMFA), given by \cite{RCR.98}
\begin{eqnarray}
Z_{\rm CMFA}&=& e^{-\beta \sum_\nu x_\nu^2/2v_\nu}Z(x)C_0(x)C_{\rm RPA}(x)
\,,\label{Zsp}
\end{eqnarray}
where $x$ is the value which minimizes the ``separable'' free energy
${\cal F}(x)=\sum_\nu x_\nu^2/2v_\nu-T\ln Z(x)$ and is determined by the
self-consistent ``Hartree'' equations
\begin{equation}x_\nu=v_\nu\langle Q^\nu\rangle_x\label{mfx}\,,\end{equation}
with $\langle Q^\nu\rangle_x=\sum_{i,k}p_{k_i}\langle k_i|Q^\nu_i|k_i\rangle$.
$C_0(x)$ accounts for the small amplitude static fluctuations and is given by
\begin{eqnarray}
C_0(x)&=&{\rm Det}[v_\nu\frac{\partial^2{\cal F}(x)}
{\partial x_{\nu}\partial x_{\nu'}}]^{-1/2}\nonumber\\
&=&{\rm Det}[\delta_{\nu\nu'}+v_\nu(R_{\nu\nu'}(x,0)-
\sum_{i,k}\langle k_i|Q^{\nu}_i|k_i\rangle\frac{\partial p_{k_i}}
{\partial x_{\nu'}})]^{-1/2}\,,\label{C0}
\end{eqnarray}
with $\partial p_{k_i}/\partial x_{\nu'}= \beta p_{k_i}\sum_{k'_i}\langle
k'_i|Q^{\nu'}_i|k'_i\rangle (\delta_{kk'}-p_{k'_i})$.

Away from critical points Eq.\ (\ref{Zsp}) can be employed right up to
$T\rightarrow 0$. Note, however, that if the solution of (\ref{mfx}) exhibits a
continuous degeneracy (due to a continuous symmetry violation by $h(x)$) the
previous approach should be applied just to the intrinsic variables (see sec.\
III). In this case the lowest RPA energy vanishes at the mean field solution
\cite{RC.97,RCR.98} but Eq.\ (\ref{Zsp}) is still applicable, as $C_{\rm
RPA}(x)$, Eq.\ (\ref{Crpax}), remains {\it finite} for
$\omega_\alpha\rightarrow 0$. Omitting $C_{\rm RPA}(x)$ and $C_0(x)$ in
(\ref{Zsp}) leads to the plain MFA, which corresponds to a separable (product)
density.

We may then employ Eqs.\ (\ref{Zcspax}) or (\ref{Zsp}) to calculate the
two-site averages $\langle O_i^\nu O_j^{\nu'}\rangle= (2/\beta)\partial \ln
Z/\partial v^{ij}_{\nu\nu'}$ required to evaluate the reduced density
$\rho_{ij}$ and hence a certain monotone or measure of the entanglement between
subsystems $i$ and $j$. If not present in the original interaction, we may in
principle add the necessary terms in $V$ and set at the end
$v^{ij}_{\nu\nu'}=0$.

\section{Application}
\subsection{Fully connected XXZ Model}
We will consider $n$ qubits or spins coupled through a full range $XXZ$ type
interaction in the presence of a transverse magnetic field $b$. The Hamiltonian
reads
\begin{subequations}\label{H}
\begin{eqnarray}
H&=&b\sum_{i=1}^ns^z_i-V\sum_{i\neq j}^n
[s^x_i s^x_{j}+s^y_{i}s^y_{j}+(1-\gamma) s^z_is^z_j]\label{Ha}\\
&=&bS_z-V[S_x^2+S_y^2+(1-\gamma) S_z^2]+E_0\,,
\label{Hb}\end{eqnarray}
\end{subequations}
where ${\bm s_i}$ denotes the spin at site $i$ (considered dimensionless),
${\bm S}=\sum_{i=1}^n {\bm s}_i$ the total spin, $\gamma$ the anisotropy and
$E_0=nV(3-\gamma)/4$. It is apparent that $H$ commutes with $S_z$ and
$S^2=S_x^2+S_y^2+S_z^2$, its eigenvalues being
\begin{equation}
E_{SM}=bM-V[S(S+1)-\gamma M^2]+E_0\,,
\end{equation}
where $M=-S,\ldots,S$ and $S=\delta,\ldots,n/2$, with $\delta=0$ $(\half)$ for
$n$ even (odd). The ensuing partition function is
\begin{equation}
Z={\rm Tr}\exp[-\beta H]=\sum_{S=\delta}^{n/2}Y(S)\sum_{M=-S}^S
 e^{-\beta E_{SM}}\,,
\label{Z}\end{equation}
where $Y(S)=(^{\;\;\;\;n}_{n/2-S})-(^{\;\;\;\;\;n}_{n/2-S-1})$, with
$Y(\frac{n}{2})=1$, is the multiplicity of states with total spin $S$ and
$S_z=M$, such that $\sum_{S=\delta}^{n/2}Y(S)(2S+1)=2^n$. In what follows we
will write
\begin{equation}V=v/n\,,\label{V}\end{equation}
such that all intensive energies $E_{SM}/n$ remain finite
for $n\rightarrow\infty$ and finite $v$.

We will analyze here the attractive case $v>0$ (and $\gamma\leq 1$), where the
ground state has maximum spin $S=n/2$ $\forall$ $b,\gamma$. If $\gamma\leq 0$,
the ground state will be fully aligned ($|M|=n/2$) $\forall$ $b\neq 0$ and no
ground state entanglement will arise, whereas if $\gamma>0$, the ground state
will exhibit as $b$ increases $n$ {\it transitions} $M\rightarrow M-1$ at
\begin{equation}b_M=\gamma v(1-2M)/n\,, \label{bm}\end{equation}
where $E_{SM}=E_{S,M-1}$, becoming fully aligned for
\begin{equation}|b|>b_c\equiv \gamma v(1-1/n)\,.\end{equation}
Thus, $b_c$ is {\it the limit field for entanglement at $T=0$}, as all ground
states with $S=n/2$ and $|M|<n/2$ are entangled (see below).

\subsection{Exact concurrence}
We will examine here the entanglement of a pair of spins $(i,j)$, which is
determined by the reduced two-qubit density $\rho_{2}\equiv \rho_{ij}={\rm
Tr}_{n-\{i,j\}}\rho(T)$. In the present system $\rho(T)$ is completely
symmetric and $\rho_{2}$ will obviously be identical for all pairs $i\neq j$.
In the standard basis of $s^z_i,s^z_j$ eigenstates, it will have the form
\begin{equation}
\rho_{2}=\left(\begin{array}{cccc}p_{+}&0&0&0\\0&p&\alpha&0\\0&\alpha
&p&0\\0&0&0&p_-\end{array}\right)\,,
 \label{rij}\end{equation}
where $p_++2p+p_-=1$ and
\[\begin{array}{ccl}
p_{\pm}&=&\langle(\half\pm s^z_i)(\half \pm s^z_{j})\rangle=
\frac{\langle S_z^2\rangle-n/4}{n(n-1)}+
\frac{1}{4}\pm \frac{\langle S_z\rangle}{n}\,,\\
\alpha&=&\langle s^+_is^-_{j}\rangle=\frac{\langle S^2\rangle-
\langle S_z^2\rangle-n/2}{n(n-1)}\,,
\end{array} \]
with $\langle O\rangle\equiv {\rm Tr}\,\rho\,O$ the thermal average and
$s_i^{\pm}=s^x_i\pm is^y_i$. Hence, $\rho_2$ is here completely determined by
the three collective averages $\langle S_z\rangle$, $\langle S_z^2\rangle$ and
$\langle S^2\rangle$, which can be directly derived from Eq.\ (\ref{Z}) as
(we set $k=1$)
\begin{equation}
\begin{array}{l}
\langle S_z\rangle=-T\frac{\partial \ln Z}{\partial b}\,,\;\;\;\;\;\;
\langle S_z^2\rangle=T^2\frac{\partial^2\ln Z}{\partial b^2}+
\langle S_z\rangle^2\\
\langle S^2\rangle=n T\frac{\partial \ln Z}{\partial v}+\gamma\langle
S_z^2\rangle+ \frac{n(3-\gamma)}{4}\end{array}\label{equd}\,.
\end{equation}
We may equivalently use $\langle S_z^2\rangle=(nT/v)\partial\ln
Z/\partial\gamma$.

As a measure of pairwise entanglement we will employ the {\it concurrence} $C$
\cite{W.98}, which for a general two component system can be defined as the
minimum, over all representations $\rho_{ij}=\sum_\nu q_\nu
|\Psi_\nu\rangle\langle \Psi_\nu|$, of $\sum_\nu q_\nu C(|\Psi^\nu\rangle)$,
with $C(|\Psi^\nu\rangle)=\sqrt{2[1-{\rm tr}(\rho^\nu_i)^2]}$ the square root
of the linear entropy of any of the subsystems \cite{RC.03}. The entanglement
of formation \cite{Be.96} is similarly defined but with $C(|\Psi^\nu\rangle)$
replaced by the standard entropy $-{\rm Tr}\rho^\nu_i\log_2 \rho^\nu_i$.

For a two qubit system, $C$ can be explicitly computed as \cite{W.98}
$C=[2\lambda-{\rm tr}R,0]_+$, where $[u]_+\equiv \half (u+|u|)/2$ and $\lambda$
is the largest eigenvalue of
$R=[\sqrt{\rho_{2}}\tilde{\rho}_{2}\sqrt{\rho_{2}}]^{1/2}$, with
$\tilde{\rho}_{2}=4^2s^y_is^y_j\rho_2^*s^y_js^y_i$ the spin flipped density.
The entanglement of formation becomes then just an increasing function of $C$
(given by  $E=-\sum_{\nu=\pm}q_\nu\log_2 q_\nu$, with
$q_{\pm}=(1\pm\sqrt{1-C^2})/2$), with $E=C=0$ (1) for a separable (maximally
entangled) pair.

In the present system we then obtain
\begin{eqnarray}
C&=&2[\,|\alpha|-\sqrt{p_+p_-}\,]_+\label{C}\\
&=&\frac{2}{n}[\frac{|\langle S^2\rangle-\langle S_z^2\rangle-
\frac{n}{2}|}{n-1}-\sqrt{(\frac{\langle
S_z^2\rangle+\frac{n(n-2)}{4}}{n-1})^2-\langle S_z\rangle^2}]_+\label{C2}
\end{eqnarray}
so that $\rho_{2}$ will be entangled if and only if $|\alpha|>\sqrt{p_+p_-}$, a
condition which directly follows from Peres criterion \cite{P.96}. In
(\ref{C2}) $2/n$ is the {\it maximum value} that can be attained by $C$ in
symmetric systems \cite{KBI.00}, reached here for $S=n/2$ and $M=\pm (n/2-1)$
(in which case $|SM\rangle$ is an $W$-state).

\begin{figure}[t]
\vspace*{-3.5cm}

\centerline{\hspace*{1.5cm}\scalebox{.85}{\includegraphics{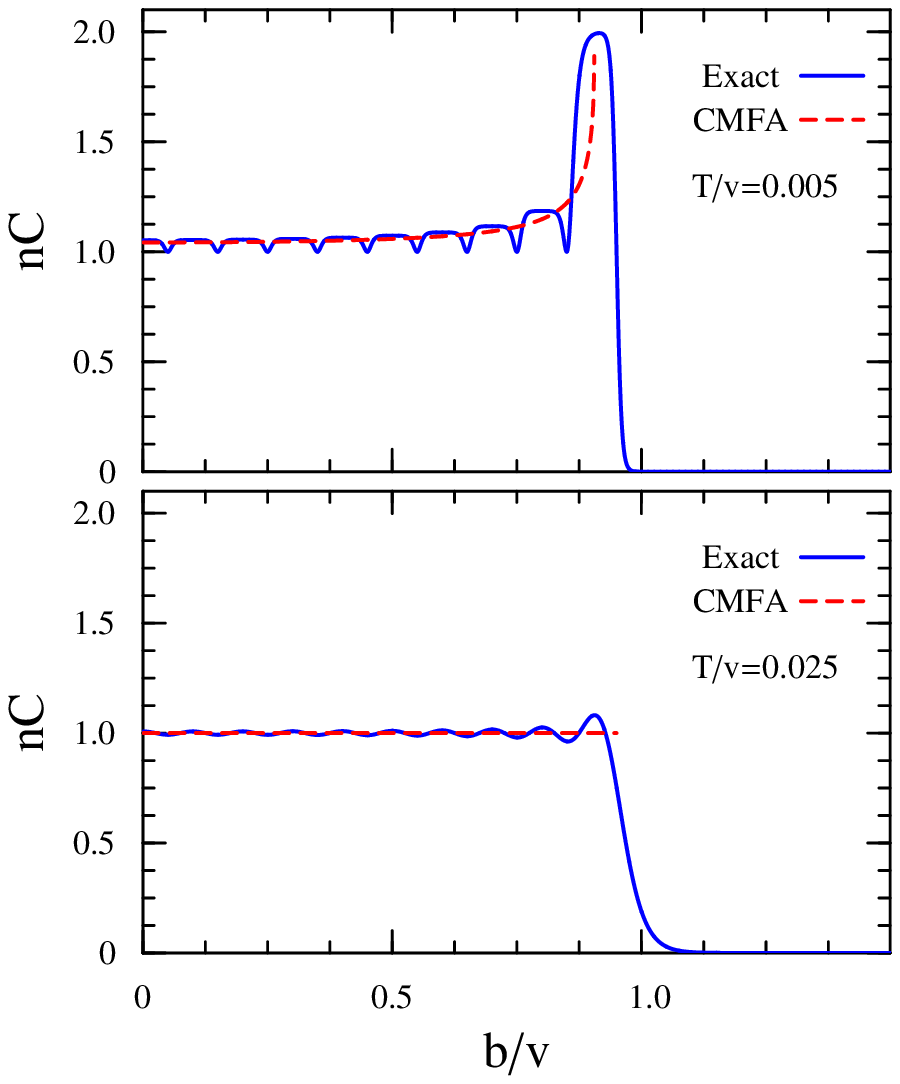}}}
 \vspace*{-12.75cm}

\caption{(Color online) Concurrence $C$ (multiplied by $n$) at low temperatures
as a function of the magnetic field $b$ for $n=20$ spins coupled through the
Hamiltonian (\ref{H}) for $\gamma=1$. The curves depict exact and CMFA results
for $T/v=0.005$ (top panel), where $C$ reaches its maximum value $2/n$ for
$b\approx b_c$, and $T/v=1/2n=0.025$ (bottom panel), where the peak at
$b\approx b_c$ is no longer prominent (Eq.\ (\ref{tl})). The $T\rightarrow 0$
behavior for any $\gamma>0$ is identical except for the rescaling $v\rightarrow
\gamma v$.}
 \label{f1}\vspace*{-0.25cm}
\end{figure}

{\it $T=0$ behavior}. Let us first briefly discuss the concurrence in the
$T\rightarrow 0$ limit, where $S^2$ and $S_z$ approach sharp values $S(S+1)$
and $M$, with $S=n/2$. Eq.\ (\ref{C2}) becomes then almost constant except for
$|M|$ close to $n/2-1$, leading, up to $O((n-1)^{-2})$, to
\begin{equation}
C\approx\frac{1}{n-1}+\frac{4m^2}{1-4m^2}\frac{1}{(n-1)^2}\,,
\end{equation}
for $m=M/n\ll 1/2$. $C$ increases stepwise from $1/(n-1)$ for $M=0$ to $2/n$
for $|M|=n/2-1$, vanishing for $|M|=n/2$ \cite{VPM.04}. The ensuing behavior of
$C$ for $n=20$ is depicted in Fig. \ref{f1}. The dips occur at the field values
(\ref{bm}) where the levels cross, in which case Eq.\ (\ref{C2}) leads to a
strictly constant lower value $C=1/n$ due to the fluctuation $\langle
S_z^2\rangle-\langle S_z\rangle^2=1/4$ at these points.

{\it Thermal behavior}. The concurrence (\ref{C2}) vanishes in symmetric states
with fixed $S$ and $M$ if $S<n/2$, with the only exception of the case
$|M|=S=n/2-1$, where $C=2/(n(n-1))$ (and $\alpha<0$). Hence, for $|b|<b_c$ we
may expect a monotonous decrease of $C$ with increasing temperature, as the
essential contribution will come from the states with $S=n/2$. The behavior for
$|b|<b_c$ will be discussed in detail in the next subsection.

Nevertheless, for $T>0$ a weak pairwise entanglement {\it also arises for}
$|b|>b_c$, i.e., when the ground state is fully separable, up to a limit
temperature $T_L$ that becomes {\it constant} for large $b$. The behavior is
thus similar to that arising with nearest neighbor $XX$ coupling \cite{CR.07}
(and in agreement with the persistence of global entanglement for large fields
in $XXZ$ models \cite{RC.05}), although here $T_L$ will decrease as $n^{-1}$
for large $n$ with the scaling (\ref{V}). To prove this result, we set $b>0$
and note that for $b-b_c\gg T$, we may just keep in $Z$ states with zero, one
and two spins up ($M=-n/2+0,1,2$) for evaluating $C$ in the lowest non-zero
order ($O(e^{-\beta b})$). This leads to
\begin{eqnarray}
C&\approx&\frac{2 e^{-\beta(b-b_c)}}{n}[1-e^{-\beta v}-
\sqrt{\frac{2n\eta}{n-1}}e^{-\beta \gamma v/n}]_+\,,\label{C4}\\
\eta&=&1-(n-1)e^{-\beta v}+\half n(n-3)e^{-2\beta v(1-1/n)}\,.\label{eta}
\end{eqnarray}
The field dependence in this limit is thus reduced to an exponential decay,
with the limit temperature  {\it $b$-independent} and determined by the root of
the bracket in (\ref{C4}) (always positive for  $T\rightarrow 0$ if
$\gamma>0$). For large $n$, $C$ is positive just for low $T\propto n^{-1}$ and
we may accurately neglect $e^{-\beta v}$ and set $\eta\approx 1$ in (\ref{C4})
(in which case it is just the result from the $S=n/2$ multiplet). This yields
\begin{equation}T_L\approx \frac{2\gamma v}{n\ln[2n/(n-1)]}
\approx \frac{2\gamma v}{n\ln 2+1}\,,\;\;b\gg b_c
\label{TL}\,.\end{equation} The maximum value reached by $C$ in this region
(attained close to $T_L$) is very small ($\propto n^{-2}e^{-n(\ln 2)(b-\gamma
v)/2}$).

\subsection{CSPA and CMFA results for the $XX$ case}
We start by describing the $XX$ case ($\gamma=1$ in (\ref{H})).
In the representation (\ref{Hb}), the CSPA, Eq.\ (\ref{Zcspax}), will lead to a
two-dimensional integral over variables $(x,y)=r(\cos\phi,\sin\phi)$ associated
with the linearized Hamiltonian $h(x,y)=bS_z-xS_x-yS_y+E_0/n$. Since
$[H,S_z]=0$, both $Z(x,y)$ and $C_{\rm RPA}(x,y)$ will be {\it independent} of
the orientation $\phi$, and the final expression can be written as
\begin{eqnarray}
&&Z_{\rm CSPA}=
\frac{n \beta}{2v}
\!\!\int_{0}^\infty\!\!\!\!rdr e^{-n\beta r^2/4v}Z(\lambda)
C_{\rm RPA}(\lambda,\omega)\,,\label{ZCSPA}\end{eqnarray}
where $\lambda=\sqrt{b^2+r^2}$ is the energy gap determined by $h(x,y)$ and
\begin{eqnarray}
&&Z(\lambda)=e^{-\beta E_0}(2\cosh{\textstyle\frac{\beta\lambda}{2}})^n\,,
\label{Zla}\\
&&C_{\rm RPA}(\lambda,\omega)=\frac{\omega\sinh(\beta\lambda/2)}
{\lambda\sinh(\beta\omega/2)}\,,\label{Crpa}\\
&&\omega=\sqrt{{\textstyle (\lambda-v\tanh\frac{\beta\lambda}{2})
(\lambda-v\frac{b^2}{\lambda^2}\tanh\frac{\beta\lambda}{2})}}\label{wr1}\,.
\end{eqnarray}
There is here a single collective RPA energy $\omega$. For temperatures lower
than the mean field critical temperature $T_c$ (see below), Eq.\ (\ref{wr1})
becomes imaginary for $r$ in an interval just below the stationary point (where
$\omega=0$), leading to the CSPA breakdown when $\omega^2<-4\pi^2T^2$. This is
first satisfied at $b=0$ and $r\approx v/2$, where $\omega\approx iv/2$,
leading to a breakdown temperature $T^*\approx v/4\pi$ ($b=0$). $T^*$ decreases
as $b$ increases, vanishing for $b>v$.

{\it CMFA.} The mean field equations (\ref{mfx}) reduce here  to
\begin{equation}
r=v\frac{r}{\lambda}\tanh{\frac{\beta\lambda}{2}},
\label{mf2}
\end{equation}
and determine the minimum  of the ``Hartree'' potential ${\cal F}(r)
=nr^2/(4v)-T\ln Z(\lambda)$. We then need to distinguish between two regimes:

a) For $|b|<v$ and $T<T_c$, where
\begin{equation}
 T_c=|b|/\ln\frac{1+|b|/v}{1-|b|/v}\,,\;\;(|b|<v) \label{Tc}
\end{equation}
the minimum of ${\cal F}(r)$ occurs at $r>0$. This solution of (\ref{mf2})
breaks the rotational symmetry around the $z$ axes and is hence continuously
degenerate ($\phi$ undetermined). In this case Eq.\ (\ref{mf2}) implies that
$\lambda$ is the root of $\lambda=v\tanh(\beta\lambda/2)$, being hence $b$-{\it
independent}, the constraint $\lambda>|b|$ leading to the critical
temperature  (\ref{Tc}) (which is a decreasing function of $|b|$, vanishing for
$|b|\rightarrow v$ and approaching $v/2$ for $b\rightarrow 0$). The gaussian
approximation (\ref{Zsp}) in the ``intrinsic'' variable $r$ leads then to
\begin{eqnarray}
Z_{\rm CMFA}&=&e^{-\frac{n\beta}{4v}(\lambda^2-b^2)}Z(\lambda)
{\textstyle\sinh\frac{\beta\lambda}{2}}
\sqrt{\textstyle\frac{4\pi n}{\beta v(1-\chi)}}\,,\label{zcmfa21}\\
\chi&=&\half\beta v/\cosh^2{\textstyle\frac{\beta\lambda}{2}}= \half\beta
v(1-{\textstyle\frac{\lambda^2}{v^2}})\,, \label{zcmfa22}
\end{eqnarray}
where the first two factors in (\ref{zcmfa21}) represent the MFA result and the
rest the RPA and SPA corrections. Note that $\omega$ {\it vanishes} at this
solution, in agreement with the broken continuous symmetry, but the RPA
correction (\ref{Crpa}) remains finite (and {\it essential}) for
$\omega\rightarrow 0$,  with $C_{\rm RPA}(\lambda,\omega)\rightarrow
\sinh(\beta\lambda/2)/(\beta\lambda/2)$.

It is apparent from (\ref{ZCSPA}) and (\ref{Zla}) that in this region the
approximation (\ref{zcmfa21}) will become increasingly accurate as $n$
increases (the $r$ fluctuation decreasing as $n^{-1}$), approaching the exact
result for $n\rightarrow\infty$.

b) For $|b|>v$ or $T>T_c$, the minimum occurs at $r=0$ (normal solution of
(\ref{mf2})). Direct application of Eq.\ (\ref{Zsp}) in the original variables
$(x,y)$ leads then to
\begin{eqnarray} Z_{\rm CMFA}&=&Z(b)\frac{\sinh(\beta b/2)}
{\sinh(\beta \omega/2)}\,,
\label{zsp0}
\end{eqnarray}
where $\omega=b-v\tanh(\beta b/2)$.

{\it Evaluation of the concurrence.} Let us first examine the CMFA concurrence
for $|b|<v$. Both $\langle S_z\rangle$ and $\langle S_z^2\rangle$ (Eq.\
(\ref{equd})) will be determined just by the MFA contribution in
(\ref{zcmfa21}), as the rest is $b$ independent, and given by
\begin{equation}
\langle S_z\rangle=-n \frac{b}{2v}\,,\;\; \langle S_z^2\rangle=\langle
S_z\rangle^2+ n\frac{T}{2v}\,,\;\;\;(T<T_c)
 \,.\label{sz1}\end{equation}
Hence, in this regime  $\langle S_z\rangle$ is {\it independent of} $T$ (it is
the value minimizing $b\langle S_z\rangle+v\langle S_z\rangle^2/n$) while the
fluctuation $\langle S_z^2\rangle-\langle S_z\rangle^2$ increases linearly with
$T$, reflecting  a gaussian distribution $p(M)\propto e^{-\beta v(M-\langle
S_z\rangle)^2/n}$. In contrast,  $\langle S^2\rangle$ is affected by all terms
in (\ref{zcmfa21}) and given by

\begin{equation}
\langle
S^2\rangle=(\frac{n\lambda}{2v})^2+\frac{n}{2}
[\frac{1-\chi[2-(1+\chi)T/v]}{(1-\chi)^2}]
 \,,\label{sx1}\end{equation}
being $b$ independent. The first term is the Hartree part. For $T\rightarrow
0$, $\lambda/v\rightarrow 1$ while $\chi\rightarrow 0$, so that (\ref{sx1})
approaches the right limit $\frac{n}{2}(\frac{n}{2}+1)$ owing to the RPA
correction.

\begin{figure}[t]
\vspace*{-3.5cm}

\centerline{\hspace*{1.5cm}\scalebox{.85}{\includegraphics{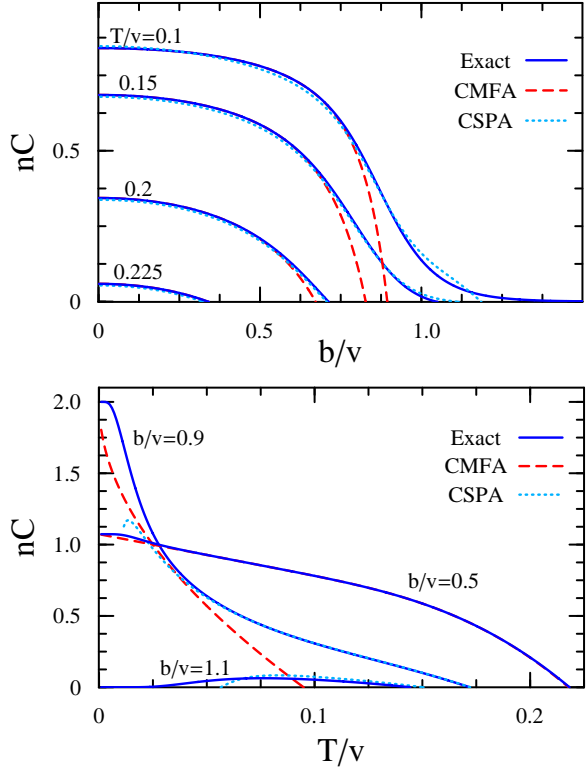}}}
 \vspace*{-11.75cm}

\caption{(Color online) Concurrence for $n=20$ spins and $\gamma=1$, as a
function of the magnetic field  at  different temperatures (top) and as a
function of temperature at different fields (bottom). Exact, CMFA and CSPA
results are depicted, that of CMFA vanishing for $b/v=1.1$ (lower panel), where
entanglement arises for $T>0$. }
 \label{f2}\vspace*{-0.25cm}
\end{figure}

The CMFA concurrence is then obtained replacing Eqs.\ (\ref{sz1})-(\ref{sx1})
in (\ref{C2}). As seen in Figs.\ (\ref{f1})-(\ref{f2}), CMFA results turn out
to be extremely accurate below the critical region, being undistinguishable
from the exact ones if $T$ is not too small. For $T\rightarrow 0$, CMFA
actually leads to the exact result but with $M$ replaced by the continuous
variable $\half nb/v$, representing then the exact $n\rightarrow\infty$ limit.
Accordingly, it does not reproduce the stepwise behavior arising for
$T\rightarrow 0$ and finite $n$, but remains close to the exact curve,
correctly predicting the peak at $b\approx b_c$ (top panel in Fig.\ \ref{f1}).

For low $T\ll T_c$, thermal effects in the CMFA will arise just from the
$S_z$ fluctuation in (\ref{sz1}), as we may still set $\chi=0$
in (\ref{sx1}). As seen in the lower panel of Fig.\ \ref{f1}, CMFA correctly
predicts the low temperature
\begin{equation}\tilde{T}=v/(2n)\,,\label{tl}\end{equation}
where the peak at $b\approx b_c$ disappears. In fact, at $T=\tilde{T}$ the CMFA
concurrence has a strictly {\it constant} value $C=1/n$ for $b <b_c$, while for
$T>\tilde{T}$ it starts to decrease with increasing field. We also note that
for $T\leq \tilde{T}$ the  CMFA result is applicable just for $b\leq b^*=
b_c-v\sqrt{1-T/\tilde{T}}/n$, becoming complex for $b>b^*$ and being maximum
just at $b=b^*$, where $C=(1+\sqrt{1-T/\tilde{T}})/n$. For $T>\tilde{T}$ it can
however be applied right up to the limit field where $C$ vanishes in the CMFA.

As seen in Fig.\ \ref{f2}, as $T$ increases beyond $\tilde{T}$, the CMFA
provides practically exact results for $C$ even for $n=20$  if $|b|\alt \half
v$, since the concurrence in this region vanishes below $T_c$.  However, the
CMFA accuracy decreases significantly if $|b|$ is close to $v$. Moreover, for
$|b|>v$ the CMFA (Eq.\ (\ref{zsp0})) is not able to reproduce the exponentially
small entanglement arising in this region. For large fields Eq.\ (\ref{zsp0})
leads to an expression similar to (\ref{C4}), i.e., $C\approx
\frac{2}{n}e^{-\beta(b-v)}[\frac{n}{n-1}(1-e^{-\beta v})
-\sqrt{\frac{2n\eta'}{n-1}}]_+$, with $\eta'\rightarrow 1$ for $T\ll v$, but
the bracket is now always negative since it lacks the last exponential factor
present in (\ref{C4}).

On the other hand, the full CSPA significantly improves CMFA results
in the wide transitional region around $b\approx v$ arising for small $n$, as
seen in Fig.\ \ref{f2}. We may also appreciate the improvement over CMFA
at field $b=0.9 v$, where the CMFA result is inaccurate for all $T$ whereas the
CSPA result is practically exact above the breakdown temperature, and also at
$b=1.1 v$, where the CMFA result vanishes while CSPA does predict the reentry
of entanglement for $T>0$, albeit above a certain onset temperature.
Nonetheless, the CSPA result cannot reproduce the exponentially small
entanglement arising for very large fields either, since it  vanishes above a
certain limit field larger than $v$.

\begin{figure}[t]
\vspace*{-1.7cm}

\centerline{\hspace*{-0.2cm}\scalebox{.675}{\includegraphics{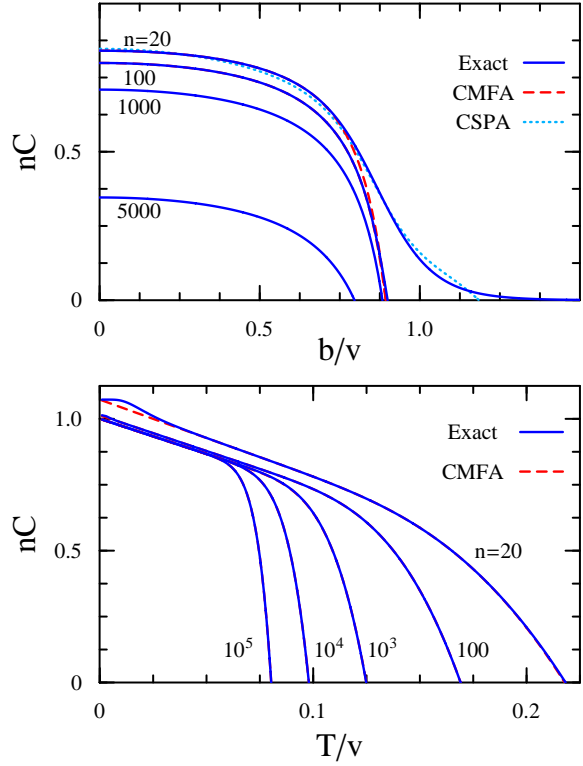}}}
 \vspace*{-8.5cm}

\caption{(Color online) Top: Concurrence as a function of the magnetic field at
$T/v=0.1$ and different values of the number $n$ of spins, for $\gamma=1$.
Exact, CMFA and CSPA results practically overlap for $n\agt 100$, the
concurrence vanishing for $n\agt 8810$. Bottom: Concurrence as function of
temperature at fixed field $b=0.5v$ and increasing values of $n$. Exact and
CMFA results are undistinguishable for $n\agt 100$.}
\label{f3}\vspace*{-0.25cm}
\end{figure}

{\it Results for large $n$.} As $n$ increases, the width of the transitional
region diminishes and the CMFA prediction for $|b|<v$ becomes increasingly
accurate, being practically exact if $n\agt 100$, as seen in Fig.\ \ref{f3}. An
expansion of the CMFA concurrence up to $O(n^{-1})$  leads to

\begin{equation}
C\approx \frac{1}{n}[\frac{1-\chi(2-(1+\chi)T/v)}{(1-\chi)^2}
-\chi\frac{T}{v}(n+1)-\frac{2T/v}{1-b^2/v^2}]_+\,,\label{C1}
\end{equation}
where the first term contains the RPA correction and provides the only positive
contribution. For $T\ll T_c$, Eq.\ (\ref{mf2}) leads to $\lambda/v\approx
1-2e^{-\beta\lambda} \approx 1-2e^{-\beta v}$, in which case $\chi\approx
2\beta v e^{-\beta v}$ and (\ref{C1}) reduces, up to order $\chi$,  to
\begin{equation} C\approx \frac{1}{n}[1-2ne^{-\beta v}
-\frac{2T/v}{1-b^2/v^2}]_+
 \label{C1a}\,.\end{equation}
Eq.\ (\ref{C1a})  provides a simple yet accurate description of $C$ for $n\agt
100$ if $|b|<v$. It implies an initial quadratic decrease with increasing
field,  and an initial $n$-independent linear decrease with increasing $T$
(Fig.\ \ref{f3}) for $2n e^{-\beta v}\ll 1$,  followed by a pronounced
$n$-dependent decrease arising from the exponential term in (\ref{C1a}) (which
represents the effect from the $S=n/2-1$ multiplet).  Note also that
at fixed $T$, entanglement will disappear for $n\agt \half e^{\beta v}(1-2T/v)$
($n\agt 8810$ in top panel of Fig.\ \ref{f3}).

\begin{figure}[t]
\vspace*{1.cm}

\centerline{\hspace*{-2.cm}\scalebox{.5}{\includegraphics{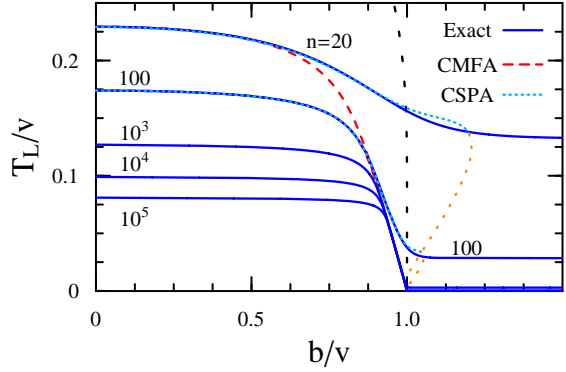}}}
 \vspace*{-0.3cm}

\caption{(Color online) Limit temperature for entanglement as a function of the
magnetic field for different values of $n$ and  $\gamma=1$. Exact, CMFA and
CSPA (for $n=20$ and $n=100$) results are depicted, undistinguishable for
$n\agt 100$. $T_L$ decreases logarithmically with $n$ for $b<v$ and as $1/n$
for $b>v$. The sparse dashed line indicates the mean field critical temperature
$T_c$.} \label{f4}\vspace*{-0.25cm}
\end{figure}

Eq.\ (\ref{C1a}) leads to a simple analytic expression for the limit field for
entanglement $b_L(T)$,
\begin{equation}b_L(T)\approx v\sqrt{1-\frac{2T/v}{1-2ne^{-\beta v}}}
 \label{blt}\,,\end{equation}
which is accurate for large $n$ and $T>T_L$ (Eq.\ (\ref{TL})), as seen in
Fig.\ \ref{f4}. The inverse of (\ref{blt}) is the limit temperature $T_L(b)$,
which is always {\it lower} than $T_c$ (Eq.\ (\ref{Tc})) and exhibits two
regimes: For $T_L(b)\ll v$ ($2ne^{-v/T_L(b)}\ll 1$),

\begin{equation}T_L(b)\approx  v-|b|\,,\end{equation}
which applies for large $n$ in a narrow field interval just before $|b|=v$,
whereas for large $n$,
\begin{equation}
T_L(b)\approx\frac{v}{\ln 2n}[1-\frac{2}{(\ln
2n)^2(1-b^2/v^2)}]\;\;\;\;(|b|<v)\,,
\end{equation}
indicating a {\it logarithmic} decrease with $n$, in contrast with the $n^{-1}$
decrease arising for $|b|>v$ (Eq.\ (\ref{TL})). For very large $n$, this yields
$T_L(b)\approx v/\ln(2n)$, independent of $b$. We also note in Fig.\ \ref{f4}
that for $n=20$ and 100, the CSPA improves the CMFA prediction of $T_L(b)$ in
the critical region, up to the field region where $T_L(b)$ becomes close to the
asymptotic value (\ref{TL}), although it also leads to a lower onset
temperature (lower sparse dotted lines).

\subsection{CSPA and CMFA in the $XXZ$  case}
In the general case ($\gamma<1$) the CSPA leads to
\begin{eqnarray}
Z_{\rm CSPA}&=&\frac{1}{4}\sqrt{\frac{n^3\beta^3}{\pi v^3(1-\gamma)}}
\!\int_{0}^\infty\!\!\!\!\!\!rdr\!\!\int_{-\infty}^{\infty}
\!\!\!\!\!\!\!\!dz\, e^{-\frac{n\beta}{4v}(r^2+\frac{z^2}{1-\gamma})}
 Z(\lambda)C_{\rm RPA}(\lambda,\omega)\label{ZCSPA3}
\end{eqnarray}
where $Z(\lambda)$ and $C_{\rm RPA}(\lambda,\omega)$ are given by Eqs.\
(\ref{Zla})-(\ref{Crpa}) with $\lambda=\sqrt{(b-z)^2+r^2}$ and
\begin{eqnarray}
\omega&=&\sqrt{{\textstyle (\lambda-v\tanh\frac{\beta\lambda}{2})
(\lambda-v(1-\gamma \frac{r^2}{\lambda^2})\tanh\frac{\beta\lambda}{2})}}
 \label{wr}\,.\end{eqnarray}
The mean field equations (\ref{mfx}) become now
\begin{equation}
r=v\frac{r}{\lambda}\tanh\frac{\beta\lambda}{2}\,,\;\;\;
z=(\gamma-1)v\frac{b-z}{\lambda}\tanh\frac{\beta\lambda}{2}\,.
\label{mf3}
\end{equation}
In the symmetry-breaking phase ($r>0$), feasible for $\gamma>0$, the solution
for $\lambda$ is then identical with that for $\gamma=1$, i.e.,
$\lambda=v\tanh(\beta\lambda/2)$, {\it independent of $b$ and $\gamma$}, in
which case (\ref{mf3}) leads to $b-z=b/\gamma$, {\it independent of $T$ and
$v$}. This implies the rescaling $b\rightarrow b/\gamma$ at the CMFA level.
This phase is then feasible for $|b|<\gamma v$ and $T<T_c$, where $T_c$ is
given by Eq.\ (\ref{Tc}) with $b\rightarrow b/\gamma$. The ensuing gaussian
approximation to both $z$ and $r$ in (\ref{ZCSPA3}) leads to the CMFA partition
function
\begin{equation}
Z_{\rm CMFA}(\gamma,b,v,T)=Z_{\rm
CMFA}(1,b/\gamma,v,T)/\sqrt{\gamma}\,,\label{Z4}
\end{equation}
where $Z_{\rm CMFA}(1,b,v,T)$ is the result (\ref{zcmfa21}).

\begin{figure}[t]
\vspace*{-1.7cm}

\centerline{\hspace*{-0.5cm}\scalebox{.675}{\includegraphics{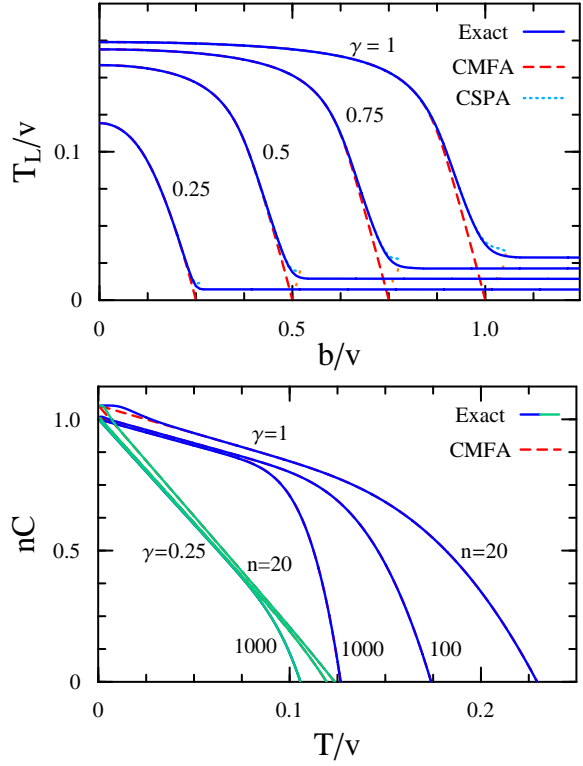}}}
 \vspace*{-8.5cm}

\caption{(Color online) Top: Limit temperature for entanglement as a function
of the magnetic field $b$ for different values of the anisotropy $\gamma$ in
(\ref{H}) and $n=100$. Exact, CSPA  and CMFA approximate results are depicted.
Bottom: Concurrence $C$ as a function of temperature at zero  field for $n=20$,
$100$ and $1000$ and two different anisotropies.}
 \label{f5}\vspace*{-0.25cm}
\end{figure}

From Eq.\ (\ref{Z4}) we obtain $\langle S_z\rangle=-\half nb/(\gamma v)$, and
$\langle S_z^2\rangle-\langle S_z\rangle^2=\half nT/(\gamma v)$ ($v\rightarrow
\gamma v$ in (\ref{sz1} and hence in (\ref{tl})), whereas $\langle
S^2\rangle$ remains unchanged (Eq.\ (\ref{sx1})). The ensuing results for $C$
exhibit the same previous features, CMFA being accurate for $|b|<\gamma v$, and
the CSPA improving the latter in the transitional region $|b|\approx \gamma v$.
This can be seen in Fig.\ (\ref{f5}), whose upper panel depicts the quenching
of the exact and approximate limit temperatures $T_L(b)$ for increasing
$\gamma$.

For large $n$ and  $T\ll T_c$, the CMFA leads now to
\begin{equation}
C\approx \frac{1}{n}
[1-2ne^{-\beta v}-\frac{2T/(\gamma v)}{1-b^2/(\gamma v)^2}]_+
\label{C1aa}\,,
\end{equation}
which generalizes Eq.\ (\ref{C1a}) and provides an accurate description for
$n\agt 100$ if $|b|<\gamma v$. For $\gamma<1$ it implies a more pronounced
initial linear decrease with increasing $T$, as seen  in  the bottom panel of
Fig.\ \ref{f5}, which for low $\gamma$ may persist up to the vanishing of $C$
even for moderate sizes ($n\approx 100$ in Fig.\ \ref{f5}). Eq.\ (\ref{C1aa})
leads to a limit field
\begin{equation}
b_L(T)\approx \gamma v\sqrt{1-\frac{2T/(\gamma v)}{1-2ne^{-\beta v}}}\,,
\end{equation}
which describes the CMFA results of Fig.\ (\ref{f5}). It implies
\[T_L(b)\approx \gamma v[1-b^2/(\gamma v)^2]\,,\;\;\;\;\;\]
for $2ne^{-v/T_L(b)}\ll 1$, a condition which may now apply $\forall$
$|b|<\gamma v$ for moderate $n$ if $\gamma$ is sufficiently low,  whereas for
sufficiently large $n$,
\[T_L(b)\approx \frac{v}{\ln (2n)}[1-\frac{2/\gamma}
 {(\ln 2n)^2[1-b^2/(\gamma v)^2]}]\,,\]
decreasing logarithmically with $n$ and becoming independent of $\gamma$ and
$b$ (for $|b|<\gamma v$) for very large $n$.

\section{Conclusions}

We have shown the feasibility of the CSPA approach for the determination of the
pairwise entanglement in composite systems at finite temperature. The method is
tractable, requiring just local diagonalizations and the evaluation of the
generalized RPA energies, and unveils the crucial role played by the RPA
correlations in the description of entanglement. It also leads to a consistent
mean-field+RPA treatment (CMFA), which, as seen in the example considered,
remains applicable and accurate in the presence of vanishing RPA energies,
arising when continuous symmetries are broken at the mean field level.

In the $XXZ$ model considered, the CMFA provides an accurate analytic
evaluation of the concurrence below the critical region ($|b|$ well below
$b_c$) even in relatively small systems, providing exact results for large $n$
if $|b|<b_c$. The full CSPA allows to extend the accuracy to the critical
region ($b\approx b_c$) in finite systems, above a low breakdown temperature,
predicting a reentry of entanglement for $T>0$ for fields above but not too far
from $b_c$, not detected at the CMFA level. Neither CMFA nor CSPA predict,
however, the exponentially small entanglement arising for large fields at low
non-zero temperatures.

The present results also reveal the rich thermal entanglement properties of the
fully connected $XXZ$ model. We have shown by means of the CMFA that the limit
temperature for pairwise entanglement decreases as $v/(\ln 2n)$ for very large
$n$ if $|b|<b_c$, whereas for $|b|>b_c$ it decreases as $\gamma v/n$, becoming
independent of $b$ for large fields. CMFA also shows that the $T=0$ peak in the
concurrence just before the transition to the aligned state at $b=b_c$
disappears at a low temperature $\tilde{T}\approx \gamma v/(2n)$. The extension
of the present approach to more complex systems, including spin chains with
general anisotropic interactions, is currently under investigation.

The authors acknowledge support from CIC (RR) and CONICET (JMM and NC) of
Argentina, respectively.

\end{document}